\documentclass[twocolumn,showpacs,preprintnumbers,amsmath,amssymb]{revtex4}

\usepackage{graphicx}
\usepackage{dcolumn}
\usepackage{bm}

\begin{document}

\preprint{APS/123-QED}

\title{Experimental \emph{(n,$\gamma$)} cross sections of the \emph{p}-process nuclei \\
$^{74}$Se and $^{84}$Sr}

\author{I. Dillmann}\email{iris.dillmann@ik.fzk.de}
 \altaffiliation[also at ]{Departement f\"ur Physik und Astronomie, Universit\"at Basel.}
\author{M. Heil}
\author{F. K\"appeler}
 \affiliation{Institut f\"ur Kernphysik, Forschungszentrum Karlsruhe, Postfach 3640, D-76021 Karlsruhe,
        Germany}
\author{T. Rauscher}
\author{F.-K. Thielemann}
 \affiliation{Departement Physik und Astronomie, Universit\"at Basel, Klingelbergstrasse 82, CH-4056 Basel,
        Switzerland}

\date{\today}

\begin{abstract}
The nucleosynthesis of elements beyond iron is dominated by the
\emph{s} and \emph{r} processes. However, a small amount of stable
isotopes on the proton-rich side cannot be made by neutron capture
and are thought to be produced by photodisintegration reactions on
existing seed nuclei in the so-called "\emph{p} process". So far
most of the \emph{p}-process reactions are not yet accessible by
experimental techniques and have to be inferred from statistical
Hauser-Feshbach model calculations. The parametrization of these
models has to be constrained by measurements on stable proton-rich
nuclei. A series of (\emph{n},$\gamma$) activation measurements,
related by detailed balance to the respective
photodisintegrations, were carried out at the Karlsruhe Van de
Graaff accelerator using the $^7$Li(\emph{p,n})$^7$Be source for
simulating a Maxwellian neutron distribution of \emph{kT}= 25 keV.
First results for the experimental (\emph{n},$\gamma$) cross
sections of the light \emph{p} nuclei $^{74}$Se and $^{84}$Sr are
reported. These experimental values were used for an extrapolation
to the Maxwellian averaged cross section at 30 keV,
$<$$\sigma$$>$$_{30}$, yielding 271$\pm$15~mb for $^{74}$Se, and
300$\pm$17~mb for the total capture cross section of $^{84}$Sr.
The partial cross section to the isomer in $^{85}$Sr was found to
be 190$\pm$10~mb.

\end{abstract}

\maketitle

\section{\label{Intro}Introduction}
Astrophysical models can explain the origin of most nuclei beyond
the iron group in a combination of processes involving neutron
captures on long (\emph{s} process) or short (\emph{r} process)
time scales \cite{bbfh57, lawi01}. However, 32 stable, proton-rich
isotopes between $^{74}$Se and $^{196}$Hg cannot be formed in that
way. Those \emph{p} nuclei are 10 to 100 times less abundant than
the \emph{s} and \emph{r} nuclei in the same mass region. They are
thought to be produced in the so-called $\gamma$ or \emph{p}
process, where proton-rich nuclei are made by sequences of
photodisintegrations and $\beta^+$ decays
\cite{woho78,woho90,raar95}. In this scenario, pre-existing seed
nuclei from the \emph{s} and \emph{r} processes are destroyed by
photodisintegration in a high-temperature environment, and
proton-rich isotopes are produced by ($\gamma$\emph{,n})
reactions. When ($\gamma$\emph{,p}) and ($\gamma,\alpha$)
reactions become comparable or faster than neutron emission within
an isotopic chain, the reaction path branches out and feeds nuclei
with lower charge number \emph{Z}. The decrease in temperature at
later stages of the \emph{p} process leads to a freeze-out via
neutron captures and mainly $\beta^+$ decays, resulting in the
typical \emph{p}-process abundance pattern with maxima at
$^{92}$Mo (\emph{N}=50) and $^{144}$Sm (\emph{N}=82).

The currently most favored astrophysical site for the \emph{p}
process is explosive burning in type II supernovae. The explosive
shock front heats the outer O/Ne shell of the progenitor star to
temperatures of 2-3 GK, sufficient for providing the required
photodisintegrations. Following the nucleosynthesis in such
astrophysical models, good agreement with the required \emph{p}
production is found, with exception of the low (\emph{A}$<$100)
and intermediate (150$\leq$\emph{A}$\leq$ 165) mass range, which
are underproduced by factors of 3-4 \cite{rahe02}. Because of
these persisting problems, alternative scenarios (such as type Ia
supernovae and X-ray bursters) have been suggested, each with
their own, inherent difficulties \cite{home91,scha98,argo03}.
Currently, however, it is not yet clear whether the observed
underproductions are due to a problem with astrophysical models or
with the nuclear physics input, i.e. reaction rates. Thus, a
necessary requirement towards a consistent understanding of the
\emph{p} process is the reduction of uncertainties in nuclear
data. By far most of the several hundreds of required
photodisintegration rates and their inverses need to be inferred
from Hauser-Feshbach statistical model calculations \cite{hafe52},
e.g. the codes NON-SMOKER \cite{rau95,rau00,rau01} and MOST
\cite{most02,most05}. Experimental data can improve the situation
in two ways, either by directly replacing predictions with
measured cross sections in the relevant energy range, or by
testing the reliability of predictions at other energies when the
relevant energy range is not experimentally accessible.

The role of (\emph{n},$\gamma$) reactions in the \emph{p} process
was underestimated for a long time, although it is obvious that
they have an influence on the final \emph{p}-process abundances.
Neutron captures compete with ($\gamma$,\emph{n}) reactions and
thus hinder the photodisintegration flux towards light nuclei,
especially at lower-\emph{Z} isotopes and even-even isotopes in
the vicinity of branching-points. The influence of a variation of
reaction rates on the final \emph{p} abundances has been studied
previously \cite{rau05,rapp04}. It turned out that the \emph{p}
abundances are very sensitive to changes of the neutron-induced
rates in the entire mass range, whereas the proton-induced and
$\alpha$-induced reaction rates are important at low and high mass
numbers, respectively.

Rayet et al. \cite{ray90} have also studied the influence of
several components in their \emph{p}-process network calculations.
Their nuclear flow schemes show that branching points occur even
at light \emph{p} nuclei, and are shifted deeper into the
proton-rich unstable region with increasing mass and temperature.
In contradiction to Woosley and Howard \cite{woho78}, who claimed
for their network calculations that (\emph{n},$\gamma$) can be
neglected except for the lightest nuclei (\emph{A}$\leq$90), Rayet
et al. also examined the influence of neutron reactions for
temperatures between \emph{T}$_9$= 2.2 and 3.2 GK by comparing
overabundance factors if (\emph{n},$\gamma$) reactions on
\emph{Z}$>$26 nuclides are considered or completely suppressed. As
a result, the overabundances were found to change by up to a
factor 100 (for $^{84}$Sr) if the (\emph{n},$\gamma$) channel was
artificially suppressed. This rather high sensitivity indicates
the need for reliable (\emph{n},$\gamma$) rates to be used in
\emph{p}-process networks.

Although recent efforts are directed to calculation or measurement
of photodisintegration cross sections and rates
\cite{movo00,vomo01,utso03a,utso03b,somo03,sovo04,rau04},
astrophysical photodisintegration rates can easily be inferred
from capture rates by detailed balance, even in theoretical work
\cite{rau00}. The stellar reaction rate
N$_A$$<$$\sigma\upsilon$$>$$^*$$_{n,\gamma}$ for the reaction
\mbox{a + \emph{n} $\rightarrow$ b + $\gamma$} is related to its
inverse rate by
\begin{eqnarray}
N_A<\sigma\upsilon>^*_{\gamma,n}= \frac{(2J_a +1)(2J_n +1)}{2J_b
+1}~\sqrt{\Big( \frac{A_a}{A_b}\Big)^3}~\times \nonumber\\
\times~\frac{G_a(T)}{G_b(T)}~exp
\Big(-\frac{Q_{n,\gamma}}{kT}\Big)~
N_A<\sigma\upsilon>^*_{n,\gamma} \label{eq:1}
\end{eqnarray}
with the Avogadro number \emph{N$_A$}, the nuclear spins \emph{J}
and masses \emph{A}, the respective temperature-dependent
partition functions \emph{G(T)} and the reaction \emph{Q} value in
the exponent. Measuring or calculating a rate in the direction of
positive \emph{Q} value ensures best numerical accuracy and
consistency between forward and backward reaction. This is
important when implementing those rates in reaction networks for
nucleosynthesis models.

Moreover, stellar cross sections and rates have to be employed for
the computation of reverse rates. In a stellar environment, nuclei
are fully thermalized with the environment, resulting in a thermal
excitation of both the target and the final nucleus. Only stellar
cross sections including the excitation in form of a stellar
enhancement factor (SEF) can be used to properly account for all
transitions when applying detailed balance. For reactions with
positive \emph{Q} values for captures, a laboratory measurement of
the capture cross section will encompass by far more of the
relevant transitions than a photodisintegration experiment, even
with the target being in the ground state \cite{vomo01}.

For the past decade there has been a continuing effort to measure
nuclear data for the \emph{p} process, both for charged particle
reactions
\cite{fuki96,saka97,somo98,somo98a,bork98,chmu99,hari01,gyso01,ozmu02,rapp02,gade03,gyfu03,hari03,rapp04}
and for neutron induced reactions
\cite{wiss96,thei98,bao00,rapp02,rapp04}. The present work
comprises the first measurement of (\emph{n},$\gamma$) cross
sections for the \emph{p}-process isotopes $^{74}$Se and $^{84}$Sr
at \emph{kT}= 25 keV, with the aim to improve the \emph{p}-process
database and to help testing theoretical predictions. Since it is
not possible to measure cross sections directly at
\emph{p}-process temperatures of \emph{kT}= 170-260 keV, we have
to perform the measurements at \emph{s}-process (and freeze-out)
temperatures of \emph{kT}= 25 keV, and then extrapolate
theoretically by means of the respective energy dependent cross
sections (see Sec.~\ref{Theory}).

The measurement of stellar (\emph{n},$\gamma$) rates requires a
"stellar" neutron source yielding neutrons with a
Maxwell-Boltzmann energy distribution. We achieve this by making
use of the $^7$Li(\emph{p,n})$^7$Be reaction. In combination with
the activation or time-of-flight technique, this offers a unique
tool for comprehensive studies of (\emph{n},$\gamma$) rates and
cross sections for astrophysics.

In Sec.~\ref{Exp}, the experimental technique and sample
charac\-teristics are outlined, followed by the description of the
data analysis in Sec.~\ref{Data}. The results are presented in
Sec.~\ref{Res}. A comparison to theory and extrapolation to higher
energies is given in Sec.~\ref{Theory}. The paper is concluded
with a summary and a short outlook in Sec.~\ref{Sum}.

\section{\label{Exp}Experimental technique}
All measurements were carried out at the Karlsruhe 3.7 MV Van de
Graaff \mbox{accelerator} using the activation technique. Neutrons
were produced with the $^7$Li(\emph{p,n})$^7$Be source by
bombarding 30 $\mu$m thick layers of metallic Li on a water-cooled
Cu backing with protons of 1912 keV, 30 keV above the reaction
threshold. The resulting quasi-stellar neutron spectrum
approximates a Maxwellian distribution for \emph{kT}= \mbox{25.0
$\pm$ 0.5 keV} \cite{raty88}. Hence, the proper stellar capture
cross section can be directly deduced from our measurement.

For the activations natural samples of selenium metal (0.89\%
$^{74}$Se) and various strontium compounds (0.56\% $^{84}$Sr) were
used. In order to verify the stoichiometry, samples of Sr(OH)$_2$
and SrF$_2$ were dried at 300$^\circ$C and 800$^\circ$C,
respectively. The powders were pressed to thin pellets, which were
enclosed in a 15 $\mu$m thick aluminium foil and sandwiched
between 10-30 $\mu$m thick gold foils of the same diameter. In
this way the neutron flux can be determined relative to the
well-known capture cross section of $^{197}$Au \cite{raty88}. The
activation measurements were carried out with the Van de Graaff
accelerator operated in DC mode with a current of
$\approx$100~$\mu$A. The mean neutron flux over the period of the
activations was $\approx$1.5$\times$10$^9$ \emph{n}/s at the
position of the samples, which were placed in close geometry to
the Li target. Throughout the irradiation the neutron flux was
recorded in intervals of 1~min using a $^6$Li-glass detector for
later correction of the number of nuclei, which decayed during the
activation (factor f$_b$ in Eq. \ref{eq:4}).

Over the course of the present measurements, a total of 17
activations (5 for Se and 12 for Sr) have been carried out with
modified experimental parameters (see Table~\ref{tab:table1}).
Five short-time activations of 3~h to 5~h were used for
determining the partial cross section of the
$^{84}$Sr(\emph{n},$\gamma$)$^{85}$Sr$^{m}$ reaction feeding the
isomer in $^{85}$Sr with a half-life of 67.6 m. The
$^{84}$Sr(\emph{n},$\gamma$)$^{85}$Sr$^{g}$ cross section to the
ground state was separately deduced from seven long-time
activations.

\begin{table}[!h]
\caption{\label{tab:table1}Activation schemes and sample
characteristics. The suffix "m" denotes short time activations for
measurements of the partial cross section to the $^{85}$Sr$^m$
isomeric state. $\Phi$$_{tot}$ gives the neutron exposure of the
sample during the activation.}
\renewcommand{\arraystretch}{1.1} 
\begin{ruledtabular}
\begin{tabular}{lccccc}
 Sample & $\varnothing$ & Mass & Atoms & t$_{a}$ & $\Phi$$_{tot}$ \\
 No. & [mm] & [mg] & $^{74}$Se or $^{84}$Sr& [h] & [neutrons] \\
\hline
\textbf{Se} & & & & & \\
se-1 & 6 & 151.8 & 1.03$\times$10$^{19}$ & 16 & 1.10$\times$10$^{14}$\\
se-2 & 10 & 200.2 & 1.36$\times$10$^{19}$ & 7 & 1.53$\times$10$^{13}$\\
se-3 & 6 & 102.2 & 6.94$\times$10$^{18}$ & 23 & 1.64$\times$10$^{14}$\\
se-4 & 10 & 207.8 & 1.41$\times$10$^{19}$ & 24 & 0.99$\times$10$^{14}$\\
se-5 & 10 & 147.8 & 1.00$\times$10$^{19}$ & 24 & 1.16$\times$10$^{14}$\\
\hline \hline
\textbf{Sr(OH)$_2$} & & & & & \\
sr-1 & 6 & 67.6 & 1.88$\times$10$^{18}$ & 23 & 1.45$\times$10$^{14}$\\
sr-2 & 10 & 161.2 & 4.47$\times$10$^{18}$ & 25 & 8.03$\times$10$^{13}$\\
sr-3\footnotemark[1] & 6 & 119.8 & 3.32$\times$10$^{18}$ & 21 & 1.59$\times$10$^{14}$\\
sr-4m & 6 & 147.5 & 4.09$\times$10$^{18}$ & 3 & 1.13$\times$10$^{13}$\\
sr-5m & 10 & 195.3 & 5.42$\times$10$^{18}$ & 3 & 2.27$\times$10$^{13}$\\
\hline
\textbf{SrF$_2$} & & & & & \\
sr-6 & 10 & 478.3 & 1.28$\times$10$^{19}$ & 24 & 1.16$\times$10$^{14}$ \\
sr-7\footnotemark[2] & 10 & 195.7 & 5.25$\times$10$^{18}$ & 43 & 1.31$\times$10$^{14}$ \\
sr-8m & 8 & 204.5 & 5.49$\times$10$^{18}$ & 4 & 3.29$\times$10$^{13}$ \\
sr-9m & 10 & 314.4 & 8.44$\times$10$^{18}$ & 5 & 2.77$\times$10$^{13}$ \\
\hline
\textbf{SrCO$_3$} & & & & & \\
sr-10 & 8 & 91.2 & 2.08$\times$10$^{18}$ & 21 & 1.64$\times$10$^{14}$ \\
sr-11 & 10 & 152.1 & 3.47$\times$10$^{18}$ & 21 & 9.30$\times$10$^{13}$ \\
sr-12m & 8 & 222.6 & 5.09$\times$10$^{18}$ & 3 & 1.91$\times$10$^{13}$ \\
\end{tabular}
\end{ruledtabular}
\footnotetext[1]{Heated at 300$^\circ$C for 4 h.}
\footnotetext[2]{Heated at 800$^\circ$C for 1 h.}
\end{table}

\begin{table*}
\caption{\label{tab:table2}Decay properties of the product nuclei.
Shown here are only the strongest transitions, which were
considered for analysis. Isotopic abundances are from Ref.
\cite{iupac}.}
\renewcommand{\arraystretch}{1.2} 
\begin{ruledtabular}
\begin{tabular}{ccccccc}
Reaction & Isot. abund. [\%] & Final state & Half life &
E$_\gamma$ [keV] & I$_\gamma$ [\%] & Ref.\\
\hline
$^{74}$Se(n,$\gamma$)$^{75}$Se & 0.89 $\pm$ 0.04 & Ground state & 119.79 $\pm$ 0.04 d & 136.0 & 58.3 $\pm$ 0.7 & \cite{nds75}\\
& & & & 264.7 & 58.9 $\pm$ 0.3 & \\
\hline $^{84}$Sr(n,$\gamma$)$^{85}$Sr & 0.56 $\pm$ 0.01 & Ground state & 64.84 $\pm$ 0.02 d & 514.0 & 95.7 $\pm$ 4.0 & \cite{nds85}\\
& & Isomer & 67.63~$\pm$~0.04 m & 151.2 (EC) & 12.9 $\pm$ 0.7 & \\
& & & & 231.9 (IT) & 84.4 $\pm$ 2.2 & \\
\hline
$^{197}$Au(n,$\gamma$)$^{198}$Au & 100 & Ground state & 2.69517 $\pm$ 0.00021 d & 411.8 & 95.58 $\pm$ 0.12 & \cite{nds198}\\
\end{tabular}
\end{ruledtabular}
\end{table*}

\section{\label{Data}Data analysis}
\subsection{General procedure}
\begin{figure*}
\includegraphics{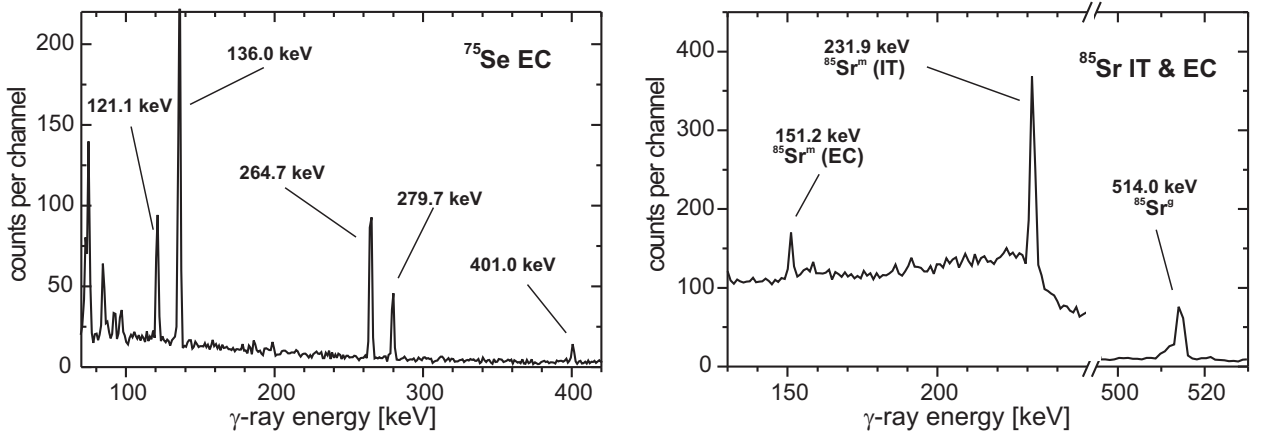}
\caption{\label{gamma}Decay spectra of the activated Se and Sr
samples. The Se spectra shows also the $\gamma$-lines at 121 keV,
280 keV and 401 keV, which were not considered for analysis.}
\end{figure*}

The induced $\gamma$-ray activities were counted after the
irradiation in a well defined geometry of 76.0$\pm$0.5~mm distance
using a shielded HPGe detector in a low background area. Energy
and efficiency calibrations have been carried out with a set of
reference $\gamma$-sources in the energy range between 60 keV and
2000 keV. Fig.~\ref{gamma} shows the $\gamma$-ray spectra of the
induced activities in the $^{74}$Se and $^{84}$Sr samples.

The total amount of activated nuclei \emph{Z} at the end of the
irradiation can be deduced from the number of events \emph{C} in a
particular $\gamma$-ray line registered in the HPGe detector
during the measuring time \emph{t$_m$} \cite{beer80}. The factor
\emph{t$_w$} corresponds to the waiting time between irradiation
and activity measurement.
\begin{eqnarray}
Z = \frac{C(t_m)} {\varepsilon_\gamma~ I_\gamma~ k_\gamma~
(1-e^{-\lambda~t_m})~e^{-\lambda~t_w}} \label{eq:2}
\end{eqnarray}
The factors $\varepsilon_\gamma$ and \emph{I$_\gamma$} account for
the HPGe efficiency and the relative $\gamma$ intensity per decay
of the respective transition (Table~\ref{tab:table2}). The factor
\emph{k$_\gamma$} introduces the correction for $\gamma$-ray
self-absorption in the sample \cite{beer80}. For disk shaped
samples with a thickness \emph{d} and $\gamma$-ray absorption
coefficients $\mu$ \cite{nist}, one obtains
\begin{eqnarray}
k_\gamma = \frac{1-e^{-\mu d}}{\mu d} \label{eq:3}.
\end{eqnarray}
The factor
\begin{eqnarray}
f_b =
\frac{\int_{0}^{t_a}\phi(t)~e^{-\lambda(t_a-t)}~dt}{\int_{0}^{t_a}\phi(t)~dt}
\label{eq:4}
\end{eqnarray}
accounts for the decay of activated nuclei during the irradiation
time \emph{t$_a$} as well as for variations in the neutron flux.
This factor can be calculated from the neutron flux history
recorded throughout the irradiation with the $^6$Li glass detector
in 91~cm distance from the target.

The number of activated nuclei \emph{Z} can also be written as
\begin{eqnarray}
Z(i) = N_i~\sigma_i~\Phi_{tot}~f_b(i) \label{eq:5},
\end{eqnarray}
where $\Phi_{tot} = \int \phi(t)dt$ is the time-integrated neutron
flux and \emph{N$_i$} the number of atoms in the sample. As our
measurements are carried out relative to $^{197}$Au as a standard,
the neutron flux $\Phi_{tot}$ cancels out:
\begin{eqnarray}
\frac{Z(i)}{Z(Au)}=
\frac{\sigma_i~N_i~f_b(i)}{\sigma_{Au}~N_{Au}~f_b(Au)} \nonumber\\
\Longleftrightarrow  \sigma_i =
\frac{Z(i)~\sigma_{Au}~N_{Au}~f_b(Au)}{Z(Au)~N_i~f_b(i)}
\label{eq:6}.
\end{eqnarray}
The reference value for the experimental $^{197}$Au cross section
in the quasi-stellar spectrum of the $^{7}$Li(\emph{p,n})$^{7}$Be
source is 586$\pm$8~mb \cite{raty88}. By averaging the induced
activities of the gold foils, one can determine the neutron flux
$\Phi$$_{tot}$ at the position of the sample and deduce the
experimental cross section $\sigma_i$ of the investigated sample
as shown in Eq.~\ref{eq:6}.

\subsection{Ground-state correction}
In the case of the activation of $^{84}$Sr, where the neutron
capture populates both, ground and isomeric state in $^{85}$Sr,
the analyzing procedure is more complicated. While the partial
cross section to the isomeric state can be easily calculated as
described above, the partial cross section to the ground state has
to be corrected for those nuclei, which decayed during activation
and measuring time already by isomeric transition.

The amount of isomer and ground state nuclei after the activation
time t$_{a}$ is described by Eq.\ref{eq:7}:
\begin{eqnarray}
Z_m(t_a)=N~\sigma_m~\Phi_{tot}~f^b_m \label{eq:7}\\
Z_g(t_a)=N~\Phi_{tot}~(\sigma_g~f^b_g~+~Y~\sigma_m~\lambda_m~g_m)
\label{eq:8}.
\end{eqnarray}
Y is the branching ratio of the isomeric transition (0.866 for
$^{85}$Sr$^{m}$), and the factor g$_{m}$ is calculated with:
\begin{eqnarray}
g_m=\frac{\int_0^{t_a} e^{-\lambda_g(t_a-t)}~dt~
\int_0^{t}\phi(t^*)~e^{-\lambda_m(t-t^*)}dt^*}{\int_0^{t_a}
\phi(t)~dt} \label{eq:9}.
\end{eqnarray}

The relation between the activity of the ground state and the
measured count rate C$_g$ can be calculated by
\begin{equation}
C_g(t_w+t_m)~=~k_\gamma
~\varepsilon_\gamma~I_\gamma\int_{t_w}^{t_w+t_m}A_g(t)~dt.
\label{eq:10}
\end{equation}
A$_{g}$(t) is further described as
\begin{equation}
\begin{split}
A_g(t)&=A_g(t_a)~e^{-\lambda_gt}~+ \\
&+~Y~\frac{\lambda_g}{\lambda_g-\lambda_m}~A_m(t_a)~(e^{-\lambda_m
t}-e^{-\lambda_g t}). \label{eq:11}
\end{split}
\end{equation}

Inserting A$_{i}$=Z$_{i}$~$\lambda_{i}$, solving the integral and
converting to Z$_g$(t$_a$) leads to
\begin{equation}
\begin{split}
Z_g(t_a)= \frac{C(t_w+t_m)}{k_\gamma
~\varepsilon_\gamma~I_\gamma~(e^{-\lambda_g t_w} - e^{-\lambda_g (t_w + t_m)})}~- \\
-~\frac{Y~\frac{\lambda_g}{\lambda_g -
\lambda_m}~Z_m(t_a)~(e^{-\lambda_m t_w} - e^{-\lambda_m
 (t_w + t_m)})}{e^{-\lambda_g t_w} - e^{-\lambda_g (t_w + t_m)}}~+ \\
+~\frac{Y~\frac{\lambda_m}{\lambda_g-\lambda_m}~Z_m(t_a)~(e^{-\lambda_gt_w}
- e^{-\lambda_g(t_w+t_m)})}{e^{-\lambda_g t_w}~-~e^{-\lambda_g
(t_w + t_m)}}. \label{eq:12}
\end{split}
\end{equation}

Thus, $\sigma_{g}$ can finally be deduced from Eq.~\ref{eq:8} by
inserting Z$_g$(t$_a$):
\begin{eqnarray}
\sigma_g= \frac{Z_g(t_a)}{N~\Phi_{tot}
~f^b_g}~-~\frac{Y~\sigma_m~\lambda_m~g_m}{f^b_g} \label{eq:13}.
\end{eqnarray}
The second term in Eq. \ref{eq:13} describes the isomeric
transition, for which the ground-state cross section has to be
corrected. For half-lives longer than the ground-state, this part
introduces major corrections, whereas for very short half-lives
this term becomes negligible.

\section{\label{Res}Results and discussion}
\subsection{General}
In an astrophysical environment with temperature \emph{T}, the
neutron spectrum corresponds to a Maxwell-Boltzmann distribution
\begin{eqnarray}
\Phi \sim E_n ~e^{-E_n /kT} \label{eq:14}.
\end{eqnarray}

The experimental neutron spectrum of the $^7$Li(\emph{p,n})$^7$Be
reaction approximates a Maxwellian distribution with \emph{kT}= 25
keV almost perfectly \cite{raty88}. But to obtain the exact
Maxwellian averaged cross section
$<$$\sigma$$>_{kT}$=$\frac{<\sigma\upsilon>}{\upsilon_T}$ for the
temperature \emph{T}, the energy-dependent cross section
$\sigma$(E) has to be folded with the experimental neutron
distribution to derive a normalization factor
NF=$\frac{\sigma}{\sigma_{exp}}$. The region beyond the resonances
(2.4~keV for $^{74}$Se and 3.5~keV for $^{84}$Sr) was then
multiplied with the normalization factor NF. The proper Maxwellian
averaged cross section as a function of thermal energy \emph{kT}
was then derived from the normalized cross section
(Fig.~\ref{dxs}) in the energy range 0.01$\leq$E$_n$$\leq$4000
keV.
\begin{equation}
\frac{<\sigma\upsilon>}{v_T}=<\sigma>_{kT}=\frac{2}{\sqrt{\pi}}
\frac{\int_{0}^{\infty} \frac{\sigma(E_n)}{NF}~E_n~e^{-E_n /kT}
~dE_n}{\int_{0}^{\infty} E_n~e^{-E_n /kT}~dE_n} \label{eq:15}.
\end{equation}
In this equation, $\frac{\sigma(E_n)}{NF}$ is the normalized
energy-dependent capture cross section and \emph{E$_n$} the
neutron energy. The factor $\upsilon_T$= $\sqrt{2kT/m}$ denotes
the most probable velocity with the reduced mass \emph{m}.

For astrophysical applications, laboratory cross sections have to
be converted to stellar cross sections involving thermally excited
targets by applying a correction factor, the so-called stellar
enhancement factor (SEF) \cite{rau00}. While there are only
comparatively few cases with low-lying nuclear states in the
\emph{s} process where the correction is important, it is to be
expected that the SEF may be larger at the much higher
\emph{p}-process temperatures. However, this is not the case for
the \emph{p} nuclei considered here, as illustrated by
Table~\ref{tab:sef}.

\begin{table}[!hb]
\caption{\label{tab:sef}Stellar enhancement factors for different
temperatures \cite{rau00}.}
\renewcommand{\arraystretch}{1.1} 
\begin{ruledtabular}
\begin{tabular}{cccc}
T [GK] & \emph{kT} [keV] & SEF($^{74}$Se) & SEF($^{84}$Sr) \\
\hline
0.3 & 26 & 1.00 & 1.00 \\
2.0 & 172 & 1.01 & 1.02 \\
2.5 & 215 & 1.02 & 1.06 \\
3.0 & 260 & 1.03 & 1.09 \\
\end{tabular}
\end{ruledtabular}
\end{table}

For $^{74}$Se and $^{84}$Sr, energy-dependent neutron capture
cross sections were available from JEFF 3.0 \cite{jeff30},
\mbox{ENDF-B VI.8} \cite{endf68}, and NON-SMOKER
\cite{rau00,rau01,nons}, whereas JENDL 3.3 \cite{jendl33} provides
only data for $^{74}$Se. The energy region between 10 eV and 2.9
keV in JEFF, ENDF-B and JENDL includes experimentally evaluated
data \cite{res} and differs only in the strength of the
resonances. The trend beyond 2.9 keV is deduced from statistical
model calculations and deviates from NON-SMOKER when the
(\emph{n,p}) channel opens at E$_n$$>$600 keV ($^{74}$Se) and
E$_n$$>$750 keV ($^{84}$Sr), respectively. In the following
sections we will use the energy dependencies of JEFF 3.0 to
determine Maxwellian averaged cross sections and compare the
results only with NON-SMOKER.

\subsection {Uncertainties}
The experimental uncertainties are summarized in
Table~\ref{tab:table3}. Since nearly every stellar neutron cross
section measurement was carried out relative to gold, the error of
1.4\% \cite{raty88} in the gold cross section cancels out.

Uncertainties between 1.4\% (for samples with 10~mm diameter) and
2.9\% (6~mm diameter) are due to an estimated sample position
uncertainty of 0.25~mm relative to the Au foils during the
activation, which affects neutron flux seen by the sample. For the
Se samples, a fairly large contribution results from the 4.5\%
error of the isotopic abundance \cite{iupac}. In the case of the
$^{84}$Sr(\emph{n},$\gamma$)$^{85}$Sr$^{g+m}$ capture cross
sections, considerable contributions come from the uncertainties
of the $\gamma$-ray intensities. The errors in the time factors
f$_b$, f$_w$=$e^{-\lambda~t_w}$ and f$_m$=$e^{-\lambda~t_m}$ are
negligible in all measurements except those of the partial cross
section to $^{85}$Sr$^{m}$ (t$_{1/2}$= 67.6 m) due to the rather
long half-lives of the product nuclei in comparison with t$_w$ and
t$_m$. The error in the masses could be neglected for all samples
except for the gold foils.

\begin{table}[!h]
\caption{\label{tab:table3}Compilation of uncertainties.}
\renewcommand{\arraystretch}{1.1} 
\begin{ruledtabular}
\begin{tabular}{lcccc}
Source of uncertainty & \multicolumn{4}{c}{Uncertainty (\%)}\\
 & $^{197}$Au & $^{74}$Se & $^{84}$Sr$\rightarrow$g & $^{84}$Sr$\rightarrow$m \\
\hline
Gold cross section & 1.4\footnotemark[1] & - & - & - \\
Isotopic abundance & - & 4.5 & 1.8 & 1.8\\
Detector efficiency & 1.5 & 1.5 & 1.5 & 1.5 \\
Divergence of n flux & - & 1.4 - 2.3 & 1.5 - 2.9 & 1.5 - 2.9 \\
Sample mass & 0.2 & - & - & -\\
$\gamma$-Ray intensity & 0.1 & 0.5/ 1.2\footnotemark[2] & 4.2\footnotemark[3] & 5.4/ 2.6\footnotemark[4]\\
$\gamma$-Ray self-absorption & - & 0.2 & 0.2 & 0.2 \\
Counting statistics & 1.0 & 0.4 - 1.6 & 3.6 - 5.3 & 0.4 - 2.0\\
Time factors f$_b$, f$_m$, f$_w$ & - & - & - & 0.2 - 1.3\\
\hline
Total uncertainty & & 5.5 - 5.7\footnotemark[5] & 6.5 - 7.9\footnotemark[5] & 4.3 - 7.1\footnotemark[5]\\
\end{tabular}
\end{ruledtabular}
\footnotetext[1]{Not included in the final uncertainty, see text.}
\footnotetext[2]{136 keV/ 265 keV.} \footnotetext[3]{514 keV}
\footnotetext[4]{151 keV/ 232 keV.} \footnotetext[5]{Incl.
uncertainty from Au.}
\end{table}

The conservatively assumed overall uncertainty for the Se
measurements is 5.7\%. The partial cross sections to the $^{85}$Sr
ground and isomeric states have uncertainties of 7.1\% and 5.3\%,
respectively, leading to a combined error of 5.6\% in the total
capture cross section. These uncertainties were also adopted for
the Maxwellian averaged cross sections, assuming that the
uncertainties of the theoretical energy dependence were negligible
for the extra\-polation to \emph{kT}= 30 keV.

\subsection {$^{74}$Se(n,$\gamma$)$^{75}$Se}
The $^{74}$Se(\emph{n},$\gamma$)$^{75}$Se reaction was analyzed
via the two strongest transitions in $^{75}$As at 136.0 keV and
264.7 keV. The results from the individual Se activations are
listed in Table \ref{tab:table4}. The capture cross section
derived with the experimental neutron distribution is
281$\pm$16~mb and was calculated as the weighted mean value of all
five activations.

\begin{table}[!h]
\caption{\label{tab:table4}Results from the Se activations.}
\renewcommand{\arraystretch}{1.1} 
\begin{ruledtabular}
\begin{tabular}{ccc}
Activation & \multicolumn{2}{c}{Cross section [mb]}\\
\textbf{$^{74}$Se(n,$\gamma$)$^{75}$Se} & (136 keV) & (265 keV) \\
\hline
se-1 & 283 $\pm$ 16 & 276 $\pm$ 16\\
se-2 & 270 $\pm$ 15 & 259 $\pm$ 14\footnotemark[1]\\
se-3 & 273 $\pm$ 16 & 265 $\pm$ 15\\
se-4 & 291 $\pm$ 16 & 287 $\pm$ 16\\
se-5 & 300 $\pm$ 17 & 284 $\pm$ 16\\
\hline
Mean cross section & \multicolumn{2}{c}{281 $\pm$ 16}\\
\end{tabular}
\end{ruledtabular}
\footnotetext[1]{Value not included in mean value.}
\end{table}

\subsection {$^{84}$Sr(n,$\gamma$)$^{85}$Sr}
In case of $^{84}$Sr, neutron captures populate both, ground and
isomeric state of $^{85}$Sr. While $^{85}$Sr$^{g}$ decays can be
identified via the 514 keV transition in $^{85}$Rb, the decay of
the isomer proceeds mainly via transitions of 232 keV and 151 keV.
The isomeric state is 239 keV above the ground state and decays
either via a 7 keV- 232 keV cascade (internal transition, 86.6\%)
or directly by electron capture (13.4\%) into the 151 keV level of
the daughter nucleus $^{85}$Rb.

The partial cross section to the isomeric state can be easily
deduced from the above mentioned transitions at 151 keV and 232
keV and yields 189$\pm$10~mb (see Table~\ref{tab:table5}). The
cross section to the ground state has to be corrected for the
internal decay of the isomer during the activation and measuring
time, and results in 112$\pm$8~mb. This leads to a total capture
cross section of 301$\pm$18~mb.

The corresponding isomeric ratio is \emph{IR}= 0.63 $\pm$ 0.04, in
perfect agreement with the value of 0.63 $\pm$ 0.06 reported for
thermal neutrons \cite{mugh81}. A NON-SMOKER estimation showed
that the isomeric ratio is almost independent of the energy
\emph{kT} in the relevant range.

\begin{table}[!h]
\caption{\label{tab:table5}Results from the Sr activations.}
\renewcommand{\arraystretch}{1.1} 
\begin{ruledtabular}
\begin{tabular}{cccc}
Activation & \multicolumn{3}{c}{cross section [mb]}\\
\textbf{$^{84}$Sr(n,$\gamma$)} & \textbf{$\rightarrow$$^{85}$Sr$^{g}$} & \multicolumn{2}{c}{\textbf{$\rightarrow$$^{85}$Sr$^{m}$}} \\
 & (514 keV) & (151 keV) & (232 keV)\\
\hline
sr-1 & 114 $\pm$ 9 &&\\
sr-2 & 124 $\pm$ 8 &&\\
sr-3 & 102 $\pm$ 8 &&\\
sr-4m & & 189 $\pm$ 13 & 194 $\pm$ 10\\
sr-5m & & 194 $\pm$ 13 & 194 $\pm$ 8\\
sr-6 & 106 $\pm$ 7 & 178 $\pm$ 12 & 189 $\pm$ 8\\
sr-7 & 107 $\pm$ 7 &&\\
sr-8m & & 186 $\pm$ 12 & 190 $\pm$ 9\\
sr-9m & & 187 $\pm$ 12 & 191 $\pm$ 8\\
sr-10 & 122 $\pm$ 8 & 192 $\pm$ 13 & 192 $\pm$ 9 \\
sr-11 & 106 $\pm$ 7  \\
sr-12m & & 178 $\pm$ 12 & 189 $\pm$ 9\\
\hline
Mean cross section & 112 $\pm$ 8 & \multicolumn{2}{c}{189 $\pm$ 10}\\
\end{tabular}
\end{ruledtabular}
\end{table}

\section{\label{Theory}Comparison with theory}
\subsection{Maxwellian cross sections for \emph{kT}= 25 keV}
\begin{figure*}
\includegraphics{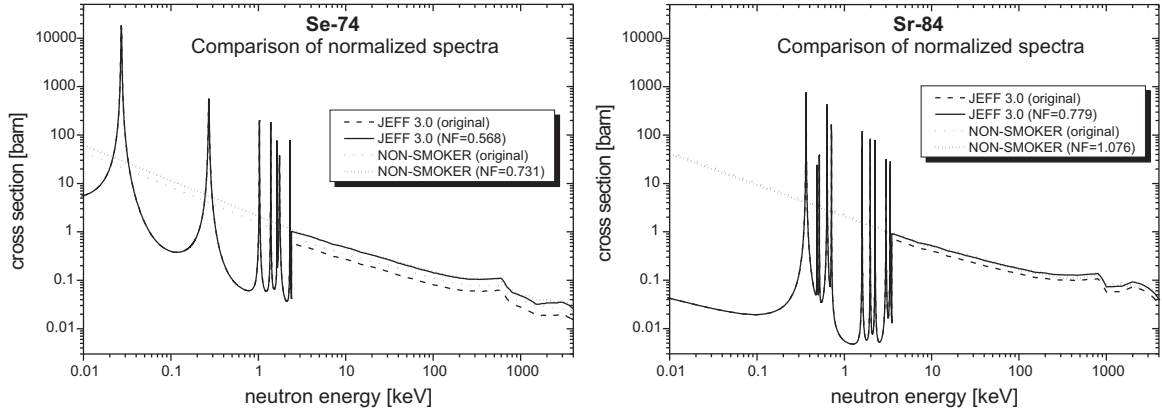}
\caption{\label{dxs}Energy-dependent cross sections $\sigma$(E)
for $^{74}$Se and $^{84}$Sr, predicted by JEFF 3.0 and NON-SMOKER.
Shown are the original data in comparison with the normalized
data. Since the resonances in both cases are experimentally, only
the curve beyond this region was normalized to our experimental
value.}
\end{figure*}

\textbf{$^{74}$Se(n,$\gamma$):} Normalization of the energy
dependent cross sections $\sigma$(E$_n$) from NON-SMOKER, JEFF
3.0, ENDF-B VI.8 and JENDL 3.3 with the experimental value of 281
mb yields normalization factors between 0.568 and 0.736 (see
Table~\ref{tab:norm}). Fig.~\ref{dxs} shows the normalized
$\sigma$(E) spectra for JEFF and NON-SMOKER in comparison with the
original spectra. The Maxwellian averaged cross section at
\emph{kT}= 25 keV deduced with the JEFF dependence is
$<$$\sigma$$>$$_{25}$= 298 mb.

\textbf{$^{84}$Sr(n,$\gamma$):} The normalization factors for
$^{84}$Sr vary between 0.774 and 1.076 (Table~\ref{tab:norm}).
With the normalized spectra of JEFF (Fig.~\ref{dxs}) the resulting
total stellar capture cross section $<$$\sigma$$>$$_{25}$(total)
is 326~mb. With our isomeric ratio of 0.63 we calculate for the
partial cross section to the isomer $^{85}$Sr$^{m}$
$<$$\sigma$$>$$_{25}$(part)= 205~mb.

\begin{table}[!h]
\caption{\label{tab:norm} Cross sections derived by folding the
experimental neutron distribution with the $\sigma$(E) data of
different libraries. NF denotes the respective normalization
factors compared to the experimental cross section.}
\renewcommand{\arraystretch}{1.1} 
\begin{ruledtabular}
\begin{tabular}{ccccc}
 & \multicolumn{2}{c}{$^{74}$Se} & \multicolumn{2}{c}{$^{84}$Sr} \\
 \cline{2-3} \cline{4-5}
 & cross section & NF & cross section & NF \\
\hline
Experiment & 281~mb & 1.000 & 301~mb\footnotemark[1] & 1.000  \\
\hline
JEFF 3.0 & 160~mb & 0.568 & 234~mb\footnotemark[1] & 0.779 \\
ENDF-B VI.8 & 160~mb & 0.568 & 233~mb\footnotemark[1] & 0.774 \\
JENDL 3.3 & 207~mb & 0.736 & - & - \\
NON-SMOKER & 206~mb & 0.731 & 324~mb\footnotemark[1] & 1.076 \\
\end{tabular}
\end{ruledtabular}
\footnotetext[1]{Total capture cross section}
\end{table}

\subsection{Extrapolation to higher temperatures}
Table \ref{tab:table7} shows the Maxwellian averaged cross
sections for different thermal energies deduced with JEFF. The
respective reaction rates were calculated from
$<$$\sigma$$>$$_{kT}$ via
\begin{eqnarray}
N_A<\sigma v>= 26445.5 <\sigma>_{kT} \sqrt{kT/m} \label{eq:17}
\end{eqnarray}
with \emph{m} being the reduced mass. The units for
$<\sigma>_{kT}$, the thermal energy \emph{kT} and the reaction
rate $N_A<\sigma v>$ are [mb], [keV] and
[mole$^{-1}$~cm$^3$~s$^{-1}$], respectively. By convention,
stellar neutron capture rates for \emph{s}-process studies are
compared at \emph{kT}= 30 keV, which corresponds also to the
freeze-out temperature of 3.5$\times$10$^8$ K. For
\emph{p}-process applications, the cross sections should be
extrapolated to the energy range, which is relevant for the
\emph{p} process, i.e. 170~keV$\leq$ \emph{kT}$\leq$ 260~keV,
corresponding to temperatures of \mbox{(2 to 3)$\times$10$^9$ K}.

For $^{74}$Se a Maxwellian averaged cross section of
$<$$\sigma$$>$$_{30}$= 271~mb is derived, in perfect agreement
with the previously estimated value of 267$\pm$25~mb from
Ref.~\cite{bao00}. The result for $^{84}$Sr is
$<$$\sigma$$>$$_{30}$(tot)= 300~mb, 18\% lower than the
368$\pm$125~mb from Ref.~\cite{bao00}. The partial cross section
to the isomer yields $<$$\sigma$$>$$_{30}$(part)= 190~mb.

\begin{figure*}
\includegraphics{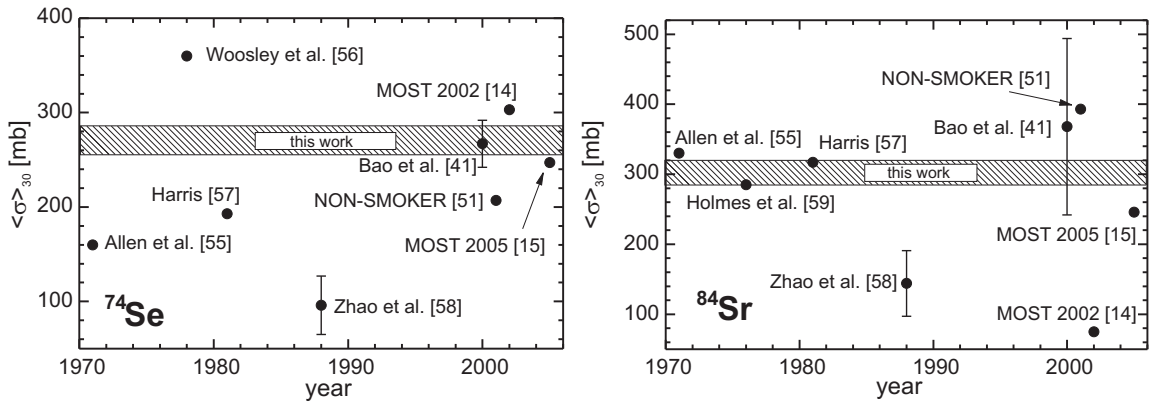}
\caption{\label{macs}Comparison of theoretically predicted
Maxwellian averaged cross sections $<$$\sigma$$>$$_{30}$ and the
experimental values (derived with the energy dependence of JEFF
3.0) for $^{74}$Se and $^{84}$Sr.}
\end{figure*}

Fig.~\ref{macs} shows theoretical predictions for the
$<$$\sigma$$>$$_{30}$ \mbox{values} of $^{74}$Se
\cite{alle71,woos78,harr81,zhao88,bao00,rau01,most02,most05} and
$^{84}$Sr
\cite{alle71,holm76,harr81,zhao88,bao00,rau01,most02,most05} in
comparison with our experimental value. In the case of $^{74}$Se
agreement is only found with the old and new MOST predictions
\cite{most02,most05} and with the normalized NON-SMOKER cross
sections in \cite{bao00}, which account for known systematic
deficiencies in the nuclear inputs of the calculation. For
$^{84}$Sr, older predictions from
Refs.~\cite{alle71,holm76,harr81} are in rather good agreement.
Not shown in this plot is a corrected prediction from the old MOST
code of 2002 \cite{gori05} of 296~mb, which was also in good
agreement with our experimental Maxwellian cross section at
\emph{kT}=30~keV. Table~\ref{tab:table6} gives a comparison
between the two Hauser-Feshbach models MOST (with the versions of
2002 \cite{most02} and 2005 \cite{most05}) and NON-SMOKER, and the
previous recommended (semi-empirical) value from
Ref.~\cite{bao00}. A full list of all 32 \emph{p} nuclei can be
found in \cite{dill05c}.

\begin{table}[!h]
\caption{\label{tab:table6}Present Maxwellian averaged cross
sections $<$$\sigma$$>$$_{30}$ at \emph{kT}= 30 keV compared to
the Hauser-Feshbach calculations MOST and NON-SMOKER and data in
\cite{bao00}.}
\begin{ruledtabular}
\begin{tabular}{cccc}
$^{74}$Se & Source & $<$$\sigma$$>$$_{30}$ [mb] & Reference \\
\hline
& NON-SMOKER & 207 & \cite{rau01}\\
& MOST 2002 & 304 & \cite{most02} \\
& MOST 2005 & 247 & \cite{most05} \\
& Bao et al. & 267 $\pm$ 25 & \cite{bao00} \\
& This work & 271 $\pm$ 15 &  \\
\hline
$^{84}$Sr & Source & $<$$\sigma$$>$$_{30}$ [mb]\footnotemark[1] & Reference \\
\hline
& NON-SMOKER & 393 & \cite{rau01}\\
& MOST 2002 & 74\footnotemark[2] & \cite{most02} \\
& MOST 2005 & 246 & \cite{most05} \\
& Bao et al. & 368 $\pm$ 125 & \cite{bao00} \\
& This work & 300 $\pm$ 17 & \\
\end{tabular}
\end{ruledtabular}
\footnotetext[1]{~Total capture cross section.}
\footnotetext[2]{~Original value. Corrected value \cite{gori05} is
296 mb for $^{84}$Sr.}
\end{table}

\begin{figure*}
\includegraphics{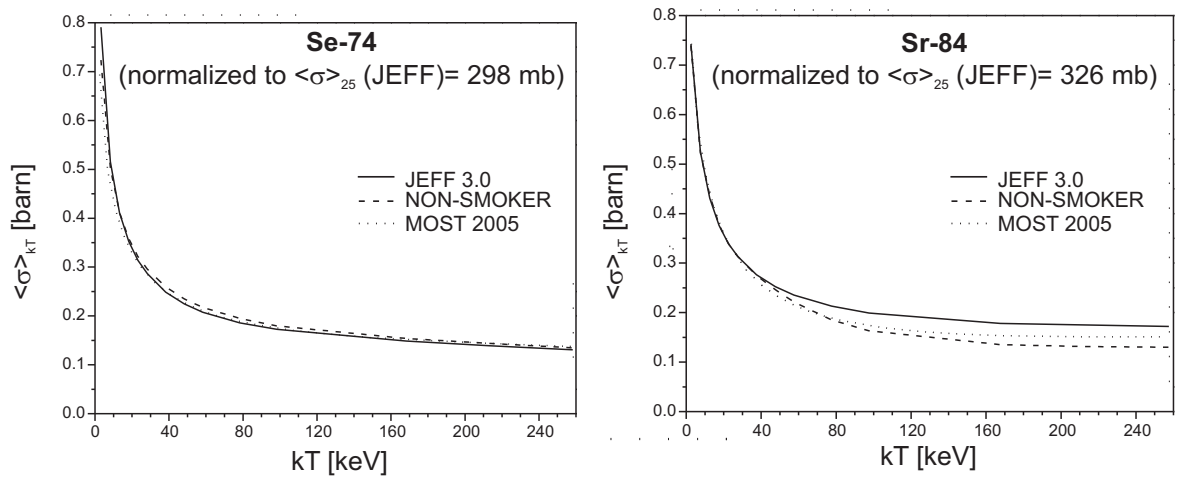}
\caption{\label{macs-kt}Temperature dependence of Maxwellian
averaged cross sections derived by normalizing different predicted
cross sections to the value deduced with JEFF 3.0 at \emph{kT}=25
keV.}
\end{figure*}

\begin{figure*}
\includegraphics{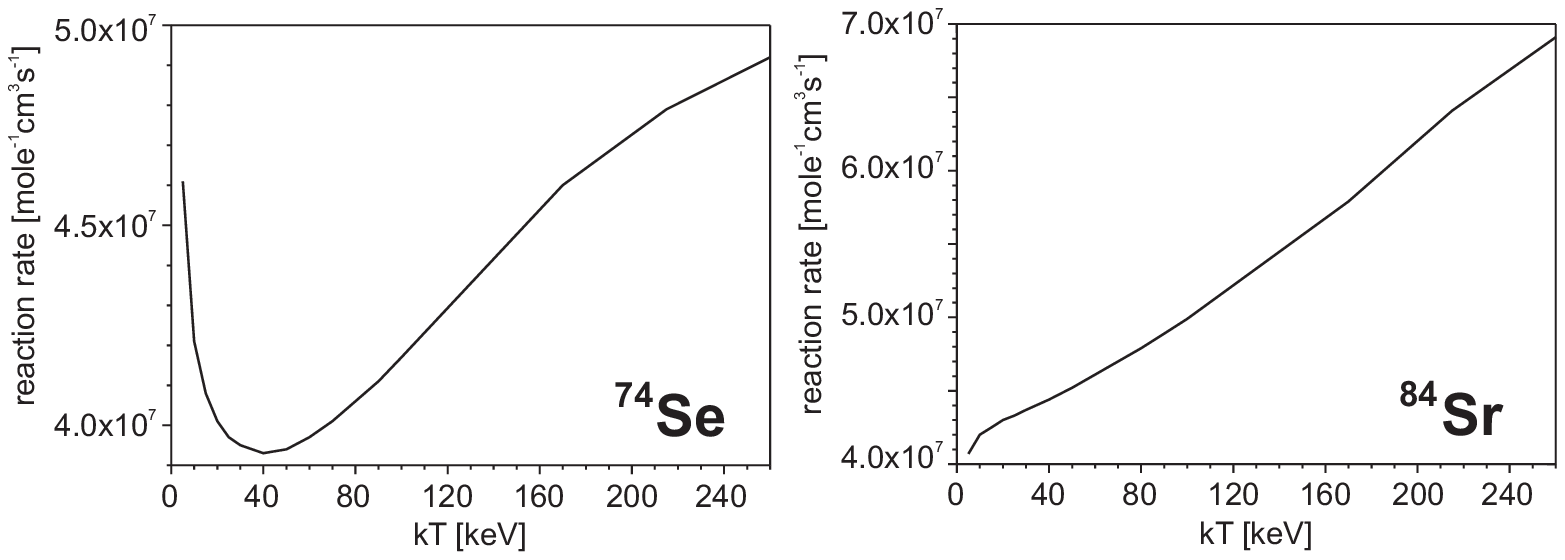}
\caption{\label{rr}Temperature trend of the reaction rates for
$^{74}$Se and $^{84}$Sr derived with the normalized
energy-dependent cross sections of JEFF 3.0 (see
Table~\ref{tab:table7}).}
\end{figure*}

Further extrapolation to temperatures between \emph{kT}= 5 keV and
260 keV (Fig.~\ref{macs-kt}) shows the different energy dependence
of the data based on the NON-SMOKER, JEFF and MOST predictions.
For this plot, the curves of NON-SMOKER and MOST were normalized
to the JEFF values at \emph{kT}= 25 keV. In the case of $^{74}$Se
the data libraries deviate at low and agree at higher energies,
whereas $^{84}$Sr exhibits an opposite trend. The results for the
Maxwellian averaged cross sections at \emph{p}-process
temperatures are $<$$\sigma$$>$$_{260}$= 115~mb for $^{74}$Se and
161~mb for $^{84}$Sr, corresponding to stellar reaction rates of
4.92$\times$10$^7$ and 6.91$\times$10$^7$
mole$^{-1}$~cm$^{3}~$s$^{-1}$, respectively. The temperature
trends of the reaction rates are shown in Fig.~\ref{rr}.

\begin{table}[!h]
\caption{\label{tab:table7}Maxwellian averaged cross sections and
reaction rates (including SEF from Table~\ref{tab:sef}) for
thermal energies between \emph{kT}= 5 keV and 260 keV derived with
the energy dependence of JEFF 3.0.}
\begin{ruledtabular}
\begin{tabular}{ccccccc}
&& \multicolumn{2}{c}{\textbf{$^{74}$Se}} && \multicolumn{2}{c}{\textbf{$^{84}$Sr}} \\
\cline{3-4} \cline{6-7}
kT && $<$$\sigma$$>$$_{kT}$ & N$_A$$<\sigma\upsilon>$ && $<$$\sigma$$>$$_{kT}$\footnotemark[1] & N$_A$$<\sigma\upsilon>$\\
$[$keV$]$ && [mb] & [mole$^{-1}$cm$^{3}$s$^{-1}$] && [mb] & [mole$^{-1}$cm$^{3}$s$^{-1}$]\\
\hline
5  && 775 & 4.61$\times$10$^7$ && 683 & 4.07$\times$10$^7$ \\
10 && 500 & 4.21$\times$10$^7$ && 499 & 4.20$\times$10$^7$ \\
15 && 395 & 4.08$\times$10$^7$ && 413 & 4.25$\times$10$^7$ \\
20 && 337 & 4.01$\times$10$^7$ && 361 & 4.30$\times$10$^7$ \\
25 && 298 & 3.97$\times$10$^7$ && 326 & 4.33$\times$10$^7$ \\
30 && 271 & 3.95$\times$10$^7$ && 300 & 4.37$\times$10$^7$ \\
40 && 233 & 3.93$\times$10$^7$ && 264 & 4.44$\times$10$^7$ \\
50 && 209 & 3.94$\times$10$^7$ && 240 & 4.52$\times$10$^7$ \\
60 && 192 & 3.97$\times$10$^7$ && 224 & 4.61$\times$10$^7$ \\
80 && 170 & 4.06$\times$10$^7$ && 201 & 4.79$\times$10$^7$ \\
100 && 157 & 4.17$\times$10$^7$ && 187 & 4.99$\times$10$^7$ \\
170 && 133 & 4.60$\times$10$^7$ && 167 & 5.79$\times$10$^7$ \\
215 && 123 & 4.79$\times$10$^7$ && 164 & 6.41$\times$10$^7$ \\
260 && 115 & 4.92$\times$10$^7$ && 161 & 6.91$\times$10$^7$ \\
\end{tabular}
\end{ruledtabular}
\footnotetext[1]{Total capture cross section}
\end{table}

\section{\label{Sum}Summary and Outlook}
We have presented the first results of an ongoing experimental
program to determine more precise \emph{p}-process reaction rates
in the mass range \emph{A}=70-140. The stellar (\emph{n},$\gamma$)
cross sections of the \emph{p} nuclei $^{74}$Se and $^{84}$Sr have
been measured for the first time, yielding values of 281~mb for
$^{74}$Se, and 112~mb for the ground and 189~mb for the isomeric
state to $^{85}$Sr with our experimental neutron spectrum. The
respective Maxwellian averaged cross sections for \emph{kT}= 30
keV were derived with the energy dependence of JEFF 3.0 and result
in $<$$\sigma$$>$$_{30}$= 271$\pm$15~mb for $^{74}$Se, and
$<$$\sigma$$>$$_{30}$(total)= 300 $\pm$17~mb for $^{84}$Sr. The
isomeric ratio IR was found to be 0.63 $\pm$ 0.04 and thus yields
a partial stellar cross section of $<$$\sigma$$>$$_{30}$(part.)=
190$\pm$10 mb.

Over the past decade, a lot of work has been devoted to measure
cross sections and reaction rates of \emph{p} nuclei, but
experimental (\emph{p},$\gamma$), ($\alpha$,$\gamma$) and
photodisintegration rates are still very scarce. The situation for
stellar (\emph{n},$\gamma$) cross sections is somewhat better, but
it should be pointed out that nearly all of the
(\emph{n},$\gamma$) measurements were performed in energy regions
relevant for the \emph{s} process (\emph{kT}= 30 keV instead of
100$<$\emph{kT}$<$260 keV for the \emph{p} process), whereas the
charged particle rates are measured close to the respective
\emph{p}-process Gamow window.

The measurements presented in this paper mark the beginning of an
extensive experimental program to determine more precise neutron
cross sections of stable \emph{p} nuclei. Within this program, we
have already finished the measurement on $^{96}$Ru \cite{rapp02},
and preliminary values are available for $^{102}$Pd, $^{120}$Te,
$^{130}$Ba, $^{132}$Ba and $^{174}$Hf \cite{dill05c}. All
available experimental information will be summarized in an
upcoming paper, including an extrapolation to the full range of
\emph{p}-process temperatures and the calculation of inverse
reaction rates by detailed balance.

\begin{acknowledgments}
We thank E. P. Knaetsch, D. Roller and W. Seith for their help and
support during the irradiations at the Van de Graaff accelerator.
This work was supported by the Swiss National Science Foundation
Grants 2024-067428.01 and 2000-105328.
\end{acknowledgments}


\end{document}